\parskip=2pt

\def\A{{\cal A}}

\def\C{{\cal C}}

\def\R{{\bf R}}
\def\I{{\cal I}}
\def \d{{\rm d}}
\def\pd{\partial}
\def\sp{\sum\nolimits^*}

\def \a{\alpha}
\def \b{\beta}
\def \om{\omega}
\def \ka{\kappa}

\def \l{\lambda}
\def \phi{\varphi}
\def \th{\theta}
\def \ep{\epsilon}
\def \s{\sigma}
\def \tr {transformation}
\def \com {constant of motion}
\def \coms {constants of motion}
\def \sy {symmetry}
\def \sys {symmetries}
\def \an {analytic}
\def \co {convergen}
\def \evl {eigenvalue}
\def \bif{bifurcat}
\def \sol{solution}
\def \e{{\rm e}}
\def \qq {\qquad}
\def \q {\quad}
\def \en {\eqno}
\def \cd {\cdot}
\def \grad {\nabla}
\def \Ker {{\rm Ker}}

\def \dst {\displaystyle}
\def \pn{\par\noindent}
\def \bs{\bigskip}
\def \en{\eqno}
\def \({\big(}
\def \){\big)}
\def \Ref{\bs\bs\pn{\bf References}\parskip=5 pt\parindent=0 pt}
\def \section#1{\bs\bs \pn {\bf #1} \bigskip}

\def\~#1{\widetilde #1}
\def\.#1{\dot #1}
\def\^#1{\widehat #1}

{\nopagenumbers
~ \vskip 3 truecm
\baselineskip .8cm
{\bf \centerline {Resonant Bifurcations}}
\vskip 2  truecm
\centerline {{\bf G. Cicogna}\footnote{$^{(*)}$}{E-mail : 
cicogna@difi.unipi.it} }
\centerline{Dipartimento di Fisica, Universit\`a di Pisa, }
\centerline{Via Buonarroti 2, Ed. B}
\centerline{I-56127, Pisa, Italy}

\vskip 2 truecm

\pn
{\it Abstract.}
\font\pet = cmr8 {\pet
\medskip\pn
We consider dynamical systems 
depending on one or more real parameters, and assuming that, 
for some ``critical'' value  of the parameters, the eigenvalues of the 
linear part are resonant, we  discuss the 
existence -- under suitable hypotheses -- of a general class of 
bifurcating solutions  in correspondence to this 
resonance. These bifurcating solutions include, as particular cases, the 
usual stationary and  Hopf bifurcations. The main idea 
is to transform the given dynamical system into normal form (in the 
sense of Poincar\'e-Dulac), and  to impose that the normalizing 
transformation is convergent, using the convergence conditions in the 
form given by A. Bruno. Some specially interesting situations, including
the cases of multiple-periodic solutions, and of degenerate eigenvalues in 
the presence of symmetry, are also discussed with some detail.}

\bigskip
{\hfill (to appear in J. Math. Anal. Appl.)}

\vfill\eject 
}

\pageno=2

~ \vskip 1 truecm

\section{1. Introduction}

In this paper we consider dynamical systems of the form
$$\.u=f(u,\lambda)\equiv A(\lambda)u+F(u,\lambda)\qq u(t)\in\R^n\qq 
\l\in\R^p\eqno(1)$$
depending on one or more real parameters $\lambda$, and, assuming that, 
for some ``critical'' value $\lambda=\lambda_0$ of the parameters, the 
matrix $A(\lambda_0)$ admits resonant eigenvalues, we want to discuss the 
existence -- under suitable hypotheses -- of a general class of 
``bifurcating solutions'' $u=u_{\lambda}(t)$ in correspondence to this 
resonance. These bifurcating solutions include, as particular cases, the 
usual stationary bifurcation and the Hopf bifurcation as well (see, e.g. 
[1]). The main idea 
is that to transform the given dynamical system into normal form (in the 
sense of Poincar\'e-Dulac [2-9]), and try to impose that the normalizing 
transformation is convergent. The imposition of convergence is 
essentially based on the application of the conditions given by Bruno [4-5], 
and leads to some prescriptions which may be fulfilled thanks to the presence 
of the parameters $\lambda$; in this way, the appearance of these ``resonant 
bifurcating'' solutions can be automatically deduced. 

Among these solutions, some attention is paid to discuss some situations 
with special interest (also from the physical point of view), including the 
cases of multiple-frequency periodic solutions, and of degenerate \evl s 
in the presence of symmetry, giving examples for each situation.


\section{2. Basic assumptions and preliminaries}

We need some preliminary notions and results.

Let $u=u(t)\in\R^n,\ \l\in\R^p$ and $f=f(u,\l)$ be a  vector-valued 
analytic function in a neighbourhood of some $u_0$ and $\l_0$ (it is not 
restrictive to choose $u_0=0$ and $\l_0=0$), such that 
$f(\l,0)=0$ for each $\l$ in a neighbourhood of $\l_0$. Let us denote 
the linear part of $f$ by $A(\l)u$ where
$$A(\l)=\nabla_u f(\l,0)\en(2)$$
and assume that, for some ``critical''  value of the parameters (we can 
assume that this value is just $\l_0=0$) the matrix 
$$A_0=A(0)$$
is semisimple, i.e. diagonalizable. 
Let us notice that in the case $A_0$ is not semisimple, then it could 
be uniquely decomposed into a sum of a semisimple and nilpotent part: 
$A_0=A^{(s)}+A^{(n)}$, and the foregoing discussion could be equally well 
applied to the semisimple part $A^{(s)}$, considering in particular normal 
forms with respect only to $A^{(s)}$. The introduction of normal 
forms with respect to a non-semisimple matrix requires a more difficult 
procedure and will not be considered here (cf. [10-11]).

Up to a linear change of coordinates (possibly after complexification of 
the space), we will then assume for convenience that the matrix 
$A_0$ is diagonal, with \evl s $\s_1,\ldots,\s_n$. The first important 
assumption is that, for the value $\l_0=0$, the \evl s
exhibit a resonance, i.e. there are some non-negative integers $m_i$ such 
that
$$\sum_{i=1}^n m_i\s_i=\s_j\qq ; \qq \sum_{i=1}^n m_i\ge 2\en(3)$$
for some index $j\in[1,\ldots,n]$.

Together with the given dynamical system (DS) (1),
we need also to consider its normal form (NF) (in the sense of 
Poincar\'e-Dulac [2-9]), in a neighbourhood of $u_0=0,\ \l_0=0$. As well 
known, the idea is that of performing a near-identity coordinate \tr
$$u\to v=u+\phi(u)\en(4)$$
in such a way that in the new coordinates $v$ the given DS takes 
its ``simplest'' form. To define this, consider 
the linear operator $\A$ \big(the ``homological operator'' [2-9]\big)
defined on the space of vector-valued functions $h(v)$ by
$$\A(h)=A_0v\cdot\grad h-A_0h \en(5)$$
Writing the NF in the form
$$\.v= g(v,\l)=A_0v+G(v,\l) \en(6)$$
the nonlinear terms $G(v,\l)$ are then defined by the property [2-9,11-14] 
$$G(v,\l)\in\Ker(\A) \en(7)$$ 
Actually, to be more precise, one should also consider, together with (1), 
the $p$ equations
$$\.\l=0$$
(as in the usual suspension procedure), and extend the homological 
operator $\A$ adding to the matrix $A_0$ the last $p$ columns and $p$ rows 
equal to zero, and similarly extend $G(v,\l)$ as a $(n+p)-$dimensional
vector-valued function with the last $p$ components equal to $0$. It is 
important indeed to notice that it is essential here to consider the $\l$ as 
independent variables; in this way, in particular, one has that
the bilinear terms in $v$ and $\l$ are included in $G(v,\l)$; see also 
Remark 1 below.

Let us now briefly recall the following important results. The proof of 
these can be found e.g. in [8,11-14].
\smallskip\pn
{\bf Lemma 1.} {\it  Given the matrix $A_0$, the most general NF is given 
by (6) with
$$G(v,\l)=\sum_i\b_i\(\l,\rho(v)\)B_iv \qq {\rm where}\qq 
B_i\in\C(A_0),\ \rho=\rho(v)\in\I_{A_0}\en(8)$$
where $\C(A_0)$ is the set of the matrices commuting with $A_0$, and 
$\I_{A_0}$ is the set of the \coms\ of the linear system
$$\.v=A_0v\en(9)$$
The sum in (8) is extended to a set of independent matrices $B_i$, 
the \coms\ $\rho(v)$ can be chosen in form of monomials (possibly fractional)
and the functions $\b_i$ are series or 
rational functions of the $v_i$ (see [13] for a detailed statement,
and below, for the cases of interest for our discussion). }
\medskip

It is now clear that the assumption (3) on the existence of some 
resonance among the \evl s of $A_0$ ensures that there are nontrivial 
\coms\ of the linear problem (9), and then nontrivial terms in the NF (8).

It is also well known that the coordinate \tr s taking the given DS into 
NF is usually performed by means of recursive techniques, and that in 
general the sequence of these \tr s is purely formal: indeed, only very 
special conditions can assure the \co ce of the normalizing \tr\ (NT).

Let us now recall the basic conditions, in the form given by Bruno, 
and called respectively Condition $\om$ and Condition A, which ensure 
this \co ce. The first condition is (see [4-5] for details)
\pn
{\sl Condition $\om$}: {\it let $\om_k=\min|(q,\s)-\s_j|$ for all 
$j=1,\ldots,n$ and all $n-$uple 
of nonnegative integers $q_i$ such that $1<\sum_{i=1}^n q_i<2^k$ and 
$(q,\s)=\sum_iq_i\s_i\ne\s_j$: then we require  }
$$\sum_{k=1}^\infty 2^{-k}\ln \big(\om_k^{-1}\big)<\infty$$
\pn
This is a actually very weak condition, devised to control the appearance 
of small 
divisors in the series of NT, and generalizes the Siegel-type condition:
$$|(q,\s)-\s_j|>\ep\ \Big(\sum_{i=1}^n q_i \Big)^{-\nu}$$
for some $\ep,\nu>0$, or the much simpler condition $|(q,\s)-\s_j|>\ep>0$, 
for all $n-$uple $q_i$ such that $(q,\s)\ne\s_j$ (see [2-5]). 
We explicitly assume from now on that this condition is always satisfied. 

The other one, instead, 
is a quite strong restriction on the form of the NF. To state this 
condition in its simplest form, let us assume that there is 
a straight line through the origin in the complex plane which contains 
all the eigenvalues $\s_i$ of $A_0$ 
. Then the condition reads
\pn
{\sl Condition A}: {\it there is a coordinate \tr\ $u\to v$ changing $f=A_0u+F$ 
to a NF $g=g(v)$ having the form
$$g(v)=A_0v+\a(v)A_0v$$
where $\a(v)$ is some scalar-valued power series \big(with $\a(0)=0$\big)}.
\smallskip\pn
In the case there is no line in the complex plane which satisfies the 
above property, then Condition A should be modified [4-5] (or even 
weakened: for instance, if there is a straight line through the origin 
such that all the $\s_i$ lie on the same side in the complex plane with 
respect to this line, then the 
eigenvalues belong to a Poincar\'e domain [2-5] and the \co ce is 
guaranteed without any other condition); but in all our applications 
below we shall assume for the sake of definiteness that the \evl s are 
either all real 
or purely 
imaginary; therefore, the above formulation of Condition A is enough to cover 
all the cases to be considered.
\smallskip\pn
{\it Remark 1.} It can be useful to point out that it is essential in the 
present approach to use the suspension procedure for the parameters $\l$, 
i.e. to consider $\l$ as additional variables. Indeed, let us consider, 
for instance, a simple standard Hopf-type $2-$dimensional \bif ion 
problem with $p=1$ parameter:
$$\.u=A_0u+\l I u+{\rm higher\ order\ terms}\qq {\rm where}\qq
A_0=\pmatrix{0 & 1 \cr -1 & 0} $$
and $I$ is the identity matrix. If $\l$ is kept fixed $\not=0$, the 
\evl s of the linear part $A_0+\l I$ are $\s=\l\pm i$ and, as a 
consequence, there are no (analytic nor fractional) \coms\ of the 
linearized problem, and the NF is trivially linear. Also, the NT would be 
convergent, as a consequence of Condition A (or -- more simply -- of the 
Poincar\'e criterion [2-5]). But no \bif ion can be found in this way. 
Instead, considering $\l$ as an independent variable, the linear part of the 
problem is $\.u=A_0u$ and now there are nontrivial \coms\ in the NF, i.e. 
the functions of $r^2=u_1^2+u_2^2$ and $\l$. \hfill$\diamondsuit$
\smallskip
Given the DS (1), it will be useful to rewrite its NF, according to 
Lemma 1, observing that obviously $A_0\in\C(A_0)$ and in view of the 
above Condition A, in the following ``splitted'' form
$$\.v=g(v,\l)=A_0v+\a\big(\l,\rho(v)\big) A_0v+
\sp_j\b_j\big(\l,\rho(v)\big)B_jv \qq\qq \rho(v)\in\I_{A_0}\en(10)$$
where (hereafter) $\sp_j$ is the sum extended to the matrices $B_j\not=A_0$ 
Following Bruno [4-5], we can then say that the \co ce 
of the NT is granted if $\sp_j\b_j\big(\l,\rho(v)\big)B_jv=0$. Clearly, 
``\co ce'' stands for ``\co ce in some neighbourhood of $u_0=0, \ \l_0=0$''.

Let us remark, incidentally, that an algorithmic implementation of the 
procedure for obtaining step by step the NF is possible (cf. e.g. [15,16]); 
we stress  however  that actually, in this paper, we shall not need any  
explicit calculation of NF's.
\hfill\eject

\section{3. The 2--dimensional case.}

For the sake of simplicity we consider first of all the case of 
2-dimensional DS, with some other simplifying assumptions; more general 
cases will be considered in subsequent sections.
\pn
{\bf Theorem 1.} {\it 
Given a 2-dimensional DS
$$\.u=A(\l)u+F(u,\l)\qq u\in\R^2,\ \l\in\R \en(11) $$
(i.e. $n=2,p=1)$, assume that for $\l_0=0$ the 
two \evl s $\s_1,\s_2$ of $A_0$ are nonzero, of opposite sign if real, 
and satisfy a resonance relation (3), which in this case can be more 
conveniently written in the form of commensurability of the two \evl s;
$${\s_1\over{s_1}}=-{\s_2\over{s_2}}=\th_0\en(12)$$
where $s_1,s_2$ are two positive, relatively prime, integers. Assume also 
that 
$$s_2{\d\over{\d\l}}A_{11}(\l)+s_1{\d\over{\d\l}}A_{22}(\l)
\Big|_{\l=0}\not=0\en(13)$$
then there is a ``resonant \bif ing'' solution $\^u=\^u_{\l}(t)$ 
of the form
$$\eqalign{\^u_1= & c^{(1)}\exp\big(s_1\th(\l)t\big)+
\sum_{\matrix{n_1=-\infty\cr n_1\not=s_1}}^{+\infty} 
c_{n_1}\exp\big(n_1\th(\l)t\big) \cr
           \^u_2= & c^{(2)}\exp\big(-s_2\th(\l)t\big)+
\sum_{\matrix{n_2=-\infty\cr n_2\not=-s_2}}^{+\infty} 
c_{n_2}\exp\big(n_2\th(\l)t\big)}\en(14)$$
where for $\l\to 0$
$$s_1\th(\l)\to \s_1\q ;\q -s_2\th(\l)\to \s_2\q ;\q c^{(1)},c^{(2)}\to 0 $$
and all terms in the two series are ``higher-order terms'', i.e.  terms 
vanishing more rapidly than the two leading terms, and the series are 
convergent in some time interval. There is also an \an\ \com\ along this 
solution, which, for small $\l$, has the form}
$$\rho(\^u)=\big(c^{(1)}\big)^{s_2}\big(c^{(2)}\big)^{s_1}+{\rm h.o.t.}$$
{\it Proof.} Consider the (possibly non-convergent) coordinate \tr\ $u\to 
v$ (4) which takes the given DS into NF (10). The assumption 
on the resonance of the \evl s $\s_i$ and Lemma 1 imply that the DS in NF is
expected to have the following form
$$\.v=g(\l,v)=A_0v+\a(\l,\rho)A_0v+\b(\l,\rho)Bv=(1+\a)A_0v+\b Bv\en(15)$$
where $\rho=\rho(v)=v_1^{s_2}v_2^{s_1}$ is a (monomial) \com\ of the linear 
problem $\.v=A_0v$,  $B$ is any diagonal matrix independent of $A_0$ 
(i.e. such that $s_2B_{11}+s_1B_{22}\not=0$), and $\l$ can be viewed as a 
\com\ of the enlarged system including $\.\l=0$. We now impose the Bruno 
Condition A (in the form  given above, indeed $\s_i$ are either 
real or purely imaginary) in order to ensure the convergence of the NT: this 
amounts to imposing
$$\b(\l,\rho)=0\en(16)$$
The expression of this function is clearly not known (unless the NF 
itself is known), but the manifold defined by $\b=0$ is \an\ [4], and the 
relevant behaviour of the first-order terms (in $\l$) of $\b(\l,\rho)$ can be 
inferred from the original DS $\.u=A(\l)u+\ldots$ , indeed the first-order 
terms (in $\l$) of $A(\l)u$, i.e. 
$A_0u+\l \big(\d A(\l)/\d\l\big)\big|_{\l=0}u$ are not changed by the 
NT; therefore
$$\a(0,0)=\b(0,0)=0$$
and
$$s_2{\d\over{\d\l}}A_{11}(\l)+s_1{\d\over{\d\l}}A_{22}(\l)\Big|_{\l=0}=
  \big(s_2B_{11}+s_1B_{22}\big){\pd\b\over{\pd\l}}\Big|_{\l=0,\rho=0}$$
having also used $s_2\s_1+s_1\s_2=0$. Assumption (13) then shows
that one can satisfy the condition $\b(\l,\rho)=0$, thanks to the 
implicit-function theorem, if $\l$ and $\rho$ are 
related by a function
$$\l=\l(\rho)\qq {\rm with}\qq \l(0)=0\en(17)$$
With $\b=0$, a solution of (15) is then
$$\^v(t)=\exp\big((1+\a)A_0t\big)\^v_0$$
or, putting $\th(\l)=\th_0(1+\a)$,
$$\eqalign{\^v_1(t)=&\^v_{10}\exp(s_1\th(\l)t)\cr
\^v_2(t)=&\^v_{20}\exp(-s_2\th(\l)t)} \en(18)$$
with the constraints
$$\big(\^v_1(t)\big)^{s_2}\big(\^v_2(t)\big)^{s_1}=
\big(\^v_{10}\big)^{s_2}\big(\^v_{20}\big)^{s_1}=\rho\in\I_{A_0}\qq
 {\rm and}\qq \l=\l(\rho) \en(19)$$
and where
$\a=\a\big(\l(\rho)\big)\to 0$ , $\th(\l)\to \th_0$ for $\l,v,\rho\to 0$.
On the other hand, in the \an\ manifold defined by 
$\b(\l,\rho)=0$ the NT is \co t and this \bif 
ing \sol\ corresponds to an \an\ \sol\ $\^u=\^u_\l(t)$ of the initial 
problem. The original coordinates $u$ are in fact related 
to the new ones $v$ by an \an\ \tr\ \big(the inverse of (4)\big)
$$u=v+\psi(v)$$ 
where $\psi$ is a power series in the $v_1,\ v_2$, and the
\co ce is granted on some neighbourhood of zero, say 
$|v_1|<R_1$, $|v_2|<R_2$. Notice now that the condition $\rho(v)={\rm 
const}$ does not ensure, if the \evl s are real (see the examples below), 
that the variables are bounded for all $t\in\R$; in this case, it will 
be sufficient to choose $\l$, together with $\^v_{10}$, $\^v_{20}$, small 
enough in order that the \co ce is granted in some time interval $T_1<t<T_2$.  
 
~ \hfill $\bullet$
\smallskip\pn
{\it Example 1.} Let the matrix elements $A_{ij}(\l)$ of $A(\l)$ in (11)
satisfy
$$ A_{11}(\l)\to 1\q ;\q A_{22}(\l)\to -2\q ;
\q A_{12}(\l)\ {\rm and}\ A_{21}(\l)\to 0
\q {\rm for}\q \l\to 0$$
Then $\s_1=1,\s_2=-2$, the \com\ of the linear DS $\.v=A_0v$ is 
$\rho=v_1^2v_2$, and, assuming $\d (2A_{11}+A_{22})/\d\l|_{\l=0}\not=0$, 
there is a resonant
\bif ing solution whose leading terms have the exponential behaviour
$$\eqalign{\^u_1= & c^{(1)}\exp\big((1+\a)t\big)+\ldots \cr
           \^u_2= & c^{(2)}\exp\big(-2(1+\a)t\big)+\ldots }$$
with $\a,c^{(1)},c^{(2)}\to 0$.
To give a more explicit case, let us assume that the first terms of the 
DS (11) are
$$\.u=\pmatrix{1+2\l & A_{12}(\l) \cr
               A_{21}(\l) & -2+\l }u + \pmatrix{u_1^3u_2 \cr u_1^2u_2^2}+
               {\rm h.o.t}$$
with $A_{12},A_{21}$ arbitrary vanishing functions of $\l$. 
In this case, the first nonlinear term is 
already in NF, with $B=I=$ Identity \big(cf. eq.(10)\big) and one 
easily gets $\th(\l)=1+\l/3+\ldots$, $\a=\l/3+\ldots$, whereas eq. (16) 
takes the form $\b={5\over 3}\l+\rho+\ldots=0$ giving then 
$$\l=-{3\over 5}\^u_1^2\^u_2+\ldots$$ 
\hfill$\diamondsuit$ 
\medskip
The next example, even if apparently similar, is actually a 
generalization of Theorem 1, indeed \evl s of the same sign will be 
involved. In this case, the \co ce of the NF would be guaranteed with no 
other condition, thanks to the Poincar\'e criterion; however, we need 
even in this case to impose that Condition A is fulfilled, in order to 
obtain a \bif ing solution of the same form as discussed so far.
\pn
{\it Example 2.} Let $A_0={\rm diag}(1,2)$, and $\d 
\(2A_{11}(\l)-A_{22}(\l)\)/\d\l|_{\l=0}\not=0$. Then now $\rho=v_1^2/v_2$ 
and the \bif ing solution is expected along a manifold of the form 
$\l \^u_2\propto \^u_1^2+\ldots$, obtained as \sol\ of the condition (16).
The discussion is now identical to the previous cases. \hfill$\diamondsuit$
\smallskip\pn
{\bf Corollary 1.} {\it (Hopf bifurcation) If $\s_1,\s_2$ are imaginary, 
$\s_{1,2}=\pm i\om_0$, then condition (13) becomes the standard 
``transversality condition''
$${\d\ {\rm Re}\ \s(\l)\over{\d\l}}\Big|_{\l=0}\not= 0$$
and the \bif ing solution is the usual Hopf \bif ion.}
\pn
{\it Proof}. The proof of Theorem 1 holds also in the case of 
imaginary \evl s. In this case, $s_1=s_2=1,\ \rho=v_1^2+v_2^2=r^2,\ 
\th(\l)=i\big(\om_0+\a(\l(r^2)\big)=i\om(\l)$, and a simple calculation 
shows indeed that 
$${\d (A_{11}+A_{22})\over{\d\l}}\Big|_{\l=0}=2{\d\ {\rm Re}\ 
\s\over{\d\l}}\Big|_{\l=0}$$
Notice that in this case the condition $\rho=r^2={\rm const}$ ensures 
$|v_i|\le r$ and then the \co ce is true for all $t\in\R$.\hfill$\bullet$

\section{4. DS with dimension $n>2$: reduction to a lower dimensional 
problem}

In this section, we will consider a general situation in which, given a 
$n-$dimensional DS with $n>2$, it is possible to reduce the problem to a 
lower dimensional case.
A quite simple but useful result is the following.

\smallskip\pn
{\bf Lemma 2.} {\it  Consider a $n-$dimensional DS ($n>2$), and assume that for 
$\l_0=0$ there are $r<n$ resonant \evl s, say $\s_1,\ldots,\s_r$, such
that no resonance relation of the following form
$$\sum_{h=1}^r m_h\s_h=\s_k \qq\qq k=r+1,\ldots,n \en(20)$$
exists. Then, the DS -- once in NF -- can be reduced to a  
$r-$dimensional problem putting for the remaining $n-r$ variables
$$\^v_k(t)\equiv 0 \qq\qq k=r+1,\ldots,n $$
If this $r-$dimensional problem admits a solution $\^v_h(t),\ 
h=1,\ldots,r$ (e.g., a \bif ing solution as in Theorem 1), then the 
original DS admits a solution in which the $n-r$ components $\^u_k(t),\
k=r+1,\ldots,n$, are ``h.o.t.'' with respect to the first $r$ components 
$\^u_h(t)$.}
\pn
{\it Proof.} Let us consider the NF variables $v_i,\ i=1,\ldots,n$, and 
let us introduce the shorthand notation $v'\equiv(v_1,\ldots,v_r)$ and 
$v''\equiv(v_{r+1},\ldots,v_n)$, and -- correspondingly -- 
$A_0=\pmatrix{A'_0 & \cr & A''_0}$ with $A'_0={\rm 
diag}(\s_1,\ldots,\s_r)$ and similarly for $A''_0$. According to Lemma 1, 
the matrices $B_i$ commuting with $A_0$ necessarily split in block form: 
$B_i=\pmatrix{B'_i & \cr & B''_i}$, therefore the DS in NF will  be
$$\eqalign { \.v'=&A'_0v'+\sum \b'_iB'_iv'=(1+\a)A'_0v'+\sp_j \b'_jB'_jv' \cr
 \.v''=&A''_0v''+\sum \b''_iB''_iv''=(1+\a)A''_0v''+\sp_j \b''_jB''_jv''}
 \en(21)$$
where  $\b'_i$ and $\b_i''$ are functions (of $\l$ and) of all the 
\coms\ $\rho(v',v'')$. Considering in particular the terms $\b''$, 
the assumption that there are no resonances of the 
form (20) excludes the occurrence of fractional \coms\ in 
$\I_{A_0}$ of the form $\b''=\rho'(v')/v_k$ for some $k=r+1,\ldots,n$, 
then $v''=0$ solves 
the second set of equations of the system in NF. Transformation into the 
initial coordinates $u$ shows the final statement. 
 
~ \hfill $\bullet$
\smallskip\pn
{\it Remark 2.}  A quite common situation, which is actually a special 
case of Lemma 2, is that the $n$ \evl s $\s_i$ are such that the only 
analytic or fractional \coms\ $\rho$ which can be constructed depend only 
on  $r$ variables $v'$, with the above notations (e.g. if 
$\s_1=2,-1,\sqrt{2},\sqrt{3}$). In this case, no resonance of the form 
(20) is clearly admitted and Lemma 2 holds true; notice in particular that 
in this case the NF (21)  becomes automatically ``triangular'' (cf. [13]). 
It is also clear that, given $n$ \evl s $\s_i$, it can happen that 
several different reductions are possible: e.g., if 
$\s_i=1,-2,\sqrt{2},-\sqrt{2},\sqrt{3}$,  Lemma 2 can allow 
a reduction of the NF  either into a 4-dimensional problem, or equally well 
into two independent 2-dimensional problems. \hfill$\diamondsuit$

\smallskip\pn
{\it Remark 3.}  We can clearly combine Theorem 1 and Lemma 2 to obtain a 
resonant \bif ing solution with $n\ge 2,\ r=2,\ p=1$. It can be noted 
that the leading terms of the resonant \bif ing solution obtained in this 
way can be characterized as the kernel of the linear operator $T$ defined 
by 
$$T={\d\over{\d t}}-A_0$$
acting on the linear space of the vectors $w\equiv(w_1,\ldots,w_n)$ where 
the $w_i$ are formal  series in powers of $\exp(\th_0t)$:
$$w_i=\sum_{m_i=-\infty}^{+\infty} c_{m_i} \e^{m_i\th_0t}\qq\qq 
i=1,\ldots,n \en(22)$$
where, using previously introduced notations (12), 
$$\th_0={\s_1\over{s_1}}=-{\s_2\over{s_2}}$$
It is easily seen, indeed, that the hypothesis on $\s_i$ is equivalent to 
the property that $T$ has precisely a $2-$dimensional kernel, generated by 
$w_1=\e^{s_2\th_0t}=\e^{\s_1t}$ and  $w_2=\e^{-s_1\th_0t}=\e^{\s_2t}$; 
see (14). \hfill$\diamondsuit$
\medskip
An especially interesting case occurs clearly when one of the \evl s is 
zero. This can be considered as a particular case of Lemma 2; we have 
indeed:
\pn
{\bf Corollary 2.} {\it (Stationary \bif ion) Assume that for $\l_0=0$ the 
matrix $A_0$ of the given DS admits just one \evl\ (say $\s_1$) equal to 
zero. Then, with the standard condition
$${\d A_{11}(\l)\over{\d\l}}\Big|_{\l=0}\not=0\en(23)$$
there is a stationary \bif ing solution of the form $\^u=\^u_1+\ldots$.}

\pn
{\it Proof}. With the notations introduced in the proof of Lemma 2 
we put here $v'=v_1,\ 
v''=(v_2,\ldots,v_n)$, and now $v_1\in\I_{A_0}$ being $\s_1=0$. Writing the
problem in the form (21), there cannot be terms of the form 
$\b''=\rho'(v_1)/v''$ (indeed, $\s_2,\ldots,\s_n\not=0$), and therefore 
$v''=0$ solves the second set of equations for $v''$. The remaining 
equation is then a $1-$dimensional equation
$$\.v_1=\sp_j\b'_j B'_j v_1\equiv \b^*(\l,v_1)v_1$$
and assumption (23) is easily seen to be equivalent to 
$\pd\b^*/\pd\l|_{\l=0}\not=0$, which ensures the existence of a stationary 
\bif ion, solving $\.v_1=0$ with $\l=\l(v_1)$, as in the usual situations. 
As in the previous cases, 
the proof is completed coming back to the original coordinates $u$.
\hfill$\bullet$
\hfill\eject

\section{5. DS with dimension $n>2$: a general result.}

We now consider a $n-$dimensional DS: according to the above Section, we 
can assume, for concreteness, that there is a resonance involving all the 
$n$ \evl s (see also Remark 4 below).
  
Before giving the main result of this Section, the following property may 
be useful (the proof is straightforward).
\smallskip\pn
{\bf Lemma 3.} {\it Given a DS in NF $\.v=g(v)=A_0v+G(v)$, the \coms\ 
$\rho\in\I_{A_0}$ of the linear part are in general not \coms\ of the 
full DS, but their time dependence can be expressed as a function only 
of the $\rho$ themselves: $(\d \rho/\d t)_g=\Phi(\rho)$, where 
$(\d /\d t)_g$ is the Lie derivative along the DS. If the DS satisfies 
Condition A, then these \coms\ $\rho$ are also \coms\ of the full DS
in NF $\.v=g(v)$.}

\smallskip\pn
{\bf Theorem 2.} {\it Consider the  DS (1) and assume that for the value 
$\l_0=0$ the \evl s $\s_i$ of $A_0$ are distinct, real 
or  purely imaginary, and satisfy a resonance 
relation (3). Assume also that $p=n-1$, i.e. that there are $n-1$ 
real parameters $\l\equiv(\l_1,\ldots,\l_{n-1})$, and finally that putting
$$a^{(i)}_k={\pd A_{ii}(\l)\over{\pd \l_k}}\Big|_{\l=0}
\qq\qq i=1,\ldots,n\ ;\qq k=1,\ldots,n-1 \en(24)$$
the $n\times n$ matrix $D$ constructed according to the following 
definition (notice that only the diagonal terms $A_{ii}(\l)$ of $A(\l)$ are 
involved) is not singular, i.e.:
$$\det D\equiv\det \pmatrix
        { \s_1 & a^{(1)}_1 & a^{(1)}_2 & \ldots & a^{(1)}_{n-1}    \cr
          \s_2  & a^{(2)}_1 & \ldots & &     \cr
                       \ldots\cr
          \s_n & a^{(n)}_1 & \ldots & & a^{(n)}_{n-1}   }\not=0 \en(25)$$
Then, there is, in a neighbourhood of $u_0=0,\ \l_0=0,\ t=0$, a \bif ing 
solution of the form
$$    \^u_i(t)=\big(\exp(\^\a(\l)A_0t)\big) \^u_{0i}(\l)+
            {\rm h.o.t.} \qq \ i=1,\ldots,n    \en(26) $$
where  $\^\a(\l)$ is some function of the $\l$'s such that 
$\^\a(\l)\to 1$ for $\l\to 0$.}

\smallskip\pn
{\it Proof.} As in the particular cases examined in the previous 
Sections, let us consider the given problem transformed into NF: 
$\.v=A_0v+G(v,\l)$ in the new coordinates $v$. Let us write $G(v,\l)$ in 
the following, more convenient form:
$$G(v,\l)=\sum_{i=1}^n \ka_i\big(\l,\rho(v)\big)K_iv$$ 
where $K_i\equiv{\rm diag}(0,\ldots,1,\ldots,0)$ is the diagonal matrix with 
$1$ at the $i-$th position, or also
$$\.v_i=\s_iv_i+\ka_i\big(\l,\rho(v)\big)v_i \qq{\rm (no\ sum\ over\ } 
i=1,\ldots,n) \qq \rho(v)\in\I_{A_0} \en(27)$$
Now recall that the functions $\rho(v)$ and $\ka(\l,\rho)$ can be 
fractional in the components $v_i$, but in such a way that each term 
$\ka_iK_iv$ is  a polynomial, so that the only admitted fractional terms  
have necessarily the form, {\it e.g.}, 
$\dst{{{v_2^{s_2}\cdot\ldots\cdot v_n^{s_n}}\over{v_1}}}$, and so on; the 
assumption that the $\s_i$ are distinct ensures that the functions $\ka$ 
cannot be of zero degree in the $v_i$ (i.e. of the form $v_2/v_1$, 
{\it e.g.}), then when $v\to 0$ all terms 
$\ka_i v_i$ vanish more rapidly than $v$ and one finds
$${\pd\over{\pd v_i}}\Big(\ka_i\big(\l,\rho(v)\big)v_i\Big)\Big|_{v=0}=
\ka_i(\l,0)\qq{\rm (no\ sum\ over\ } i)$$
Then,  eq. (27) can be written, at the lowest-order 
$$\.v_i=\s_iv_i+\ka_i(\l,0)v_i+\ldots=\s_iv_i+\sum_{k=1}^{n-1} 
q_{ik}\l_kv_i+\ldots +\ldots \en(28)$$
where $q_{ik}$ are the elements of a constant matrix with $n$ rows 
and $(n-1)$ columns. On the other hand, considering the original DS 
$$\.u=A(\l)u+\ldots$$
its diagonal bilinear terms (in $u$ and $\l$) $a_k^{(i)}\l_ku_i$ 
\big(using definition (24)\big) are just NF terms and  are not changed 
by the normalizing procedure; therefore, $a_k^{(i)}=q_{ik}$.
Now according to Condition A, the NF is \co t (or better: is obtained by 
a \co t NT) if one can rewrite (27) in the splitted form as in (10):
$$\.v=A_0v+\a\(\l,\rho(v)\)A_0v+\sp_j\b_j\(\l,\rho(v)\)B_jv 
\en(29)  $$
where $\a,\b_j$ are suitable combinations of the $\ka_i$, and 
can satisfy the $(n-1)$ conditions 
$$\b_j(\l,\rho)=0 \en(30)$$
In fact, the hypothesis (25) ensures precisely that one is able to do this  
and also to satisfy $\b_j(\l,\rho)=0$, by means of the implicit-function 
theorem, giving some $(n-1)$ relations (here the $\rho$ 
are considered as independent variables)
$$\l_j=\l_j(\rho) \en(31)$$
Once these $n-1$ conditions are satisfied, i.e. on the manifold defined 
just by (31) (which is an \an\ manifold, see [4]), the \co ce of the NT 
taking the initial DS into 
$$\.v=A_0v+\a(\l)A_0v=\^\a(\l)A_0v\en(32)$$
is granted. This DS can be easily solved, giving 
$$\^v(t)=\exp\big((\^\a(\l)A_0t)\big)\^v_0\en(33)$$
with the $n-1$ relations
$$\l_j=\l_j\big(\rho(\^v)\big)\qq\qq 
\rho=\rho\big(\^v(t)\big)=\rho(\^v_0)\in\I_{A_0}\en(34)$$
The desired result is then obtained, with similar remarks as in the proof 
of Theorem 1, coming  
back to the initial coordinates by means of the inverse (\co t) \tr\
$v\to u=v+\psi(v)$, where $\psi(v)$ are series of monomials of the $v_i\ 
(i=1,\ldots,n)$. \hfill$\bullet$
\medskip
It is immediately seen that, in particular, condition (13) of Theorem 1 is 
nothing but a special case of (25). As a generalization of Corollary 1, the 
case of purely imaginary \evl s is particularly interesting, because it 
corresponds to the case of coupled oscillators with multiple frequencies 
and gives, in the above hypotheses, the existence of multiple-periodic 
\bif ing solutions. We have indeed [17]:
\smallskip\pn
{\bf Corollary 3.} {\it With the same notations as before, let $n=r=4$ 
and $\s_1=-\s_2=i,\ \s_3=-\s_4=mi$ (with $m=2,3,\ldots$): then, with 
$\l\equiv(\l_1,\l_2,\l_3)\in\R^3$, and $\det D\not=0$, there is a  
double-periodic \bif ing solution preserving the frequency resonance }
$$\om_1:\om_2=1:m$$
\smallskip\pn
{\it Example 3.} Consider a 4-dimensional DS, with $u\in\R^4$, $\l\in\R^3$, 
describing two coupled oscillators with unperturbed frequencies 
$\om_1=1,\ \om_2=2$: 
$$\.u=A(\l)u+F(u,\l) \qq\qq {\rm where} \qq
   A(\l)=\pmatrix{\l_1+\l_3 & -(1+\l_2) & \l_1 & 0 \cr
                 1+\l_2 &     \l_1-\l_3 &   0  & \l_1 \cr
                 \l_2 & 0 & \l_3 & -2 \cr
                 0 & -\l_2 & 2 & \l_3}$$
For $\l=0$, the \evl s of $A_0$ are $\pm i,\pm 2i$, 
and it is easily seen that condition (25) is satisfied (the above 
procedure and notations can be extended without difficulty to the complex 
space). The NF will have the form, denoting by 
$w_a=v_1+iv_2,\ w_b=v_3+iv_4$ the new coordinates, in complex form
$$\eqalign
{\.w_a=&\ iw_a+\(\l_1+\b^{(1)}_1(\l,\rho)+i(\l_2+\b^{(1)}_2(\l,\rho))\)w_a
                                              \cr
\.w_b=&\ 2iw_b+\(\l_3+\b^{(1)}_3(\l,\rho)+i\b^{(1)}_4(\l,\rho)\)w_b}$$
where we have put 
$$\b_i(\l,\rho)=\b_i(\l,0)+\b_i^{(1)}(\l,\rho)$$
and the $\b_i$ are real functions of the three (functionally independent) 
\coms\ 
$$\rho_1=|w_a|^2=v_1^2+v_2^2\equiv r^2_a \ ;\ 
\rho_2=|w_b|^2=v_3^2+v_4^2\equiv r^2_b \  ;\ 
\rho_3=(w_a^2\overline{ w_b }+{\rm c.c.})\equiv 2r^2_ar_b\cos 2\phi $$
where $\phi$ is the time phase-shift between the two components $w_a$ and 
$w_b$. Notice that fractional \coms\ $\rho(w)$ may appear in this problem, 
e.g. $w_a^2/w_b$ or $\overline{ w_a }w_b/w_a$, etc. (which are 
functionally -- but not polynomially -- dependent on the three above), 
but this would not alter the result, as shown in the proof of the theorem.
The above NF  can be trasformed into the splitted form (29)
$$ \eqalign
{\.w_a=&\ iw_a+i\a\big(\l,\rho(w)\big)w_a+ \b_a\big(\l,\rho(w)\big)w_a
                                              \cr
\.w_b=&\ 2iw_b+2i\a\big(\l,\rho(w)\big)w_b+ \b_b\big(\l,\rho(w)\big)w_b}$$
with
$$\eqalign{ \a=&-\l_2-\b^{(1)}_2+\b^{(1)}_4 \cr
\b_a=&\l_1+\b^{(1)}_1+i\(2\l_2+2\b^{(1)}_2-\b^{(1)}_4\)  \ ;\ 
\b_b=\l_3+\b^{(1)}_3+i\(2\l_2+2\b^{(1)}_2-\b^{(1)}_4\) }$$
One can impose the \co ce of the NT solving for $\l_i=\l_i(\rho)$ the 
conditions $\b_a=\b_b=0$, which actually give three real conditions, 
and obtain a \bif ing double-periodic solution, with frequencies 
$$\om_1=1+\a(\l) \qq {\rm and}\qq \om_2=2\big(1+\a(\l)\)$$ 
\pn
Just to give a concrete example, let us imagine that the NF is such that
$$\b^{(1)}_1=-\rho_1\ ;\ \b^{(1)}_2=0\ ;\ \b^{(1)}_3=-\rho_2\ ;\  
\b^{(1)}_4=\rho_3$$
then the leading terms of the solution, in the original real variables 
$u_i$,  are
$$\^u_1=r_a \cos\om t\ ;\ \^u_2=r_a \sin\om t\ ;\ 
\^u_3=r_b \cos2\om(t+\phi)\ ;\ \^u_4=r_b \sin2\om(t+\phi)$$
with the constraints
$$\l_1=r_a^2\ ;\ \l_3=r^2_b\ ;\ \l_2=r_a^2r_b\cos2\phi\ ;\ \om=1+\l_2\ $$
producing (see especially the role of $\l_2$) a sort of  
amplitude--phase--frequency locking in the solution. \hfill$\diamondsuit$
\smallskip\pn
{\it Remark 4.}  As already remarked, it can happen that, among the $n$ 
resonant \evl s $\s_i$, as considered in Theorem 2, one can find some 
$r<n$ \evl s $\s_h$ in such a way that the assumption  of Lemma 2 is 
satisfied, and therefore the problem can be reduced -- as explained in 
the above Section -- to a $r-$dimensional problem. In this case, if one 
can also find $r-1$ parameters $\l_h$ in such a way that the 
corresponding $r\times r$ matrix $D$ is not singular, then the existence 
of another bifurcating solution is ensured by the same Theorem 2 (see the 
end of the final example 6 for a -- quite simple -- case). 
\hfill$\diamondsuit$

\section{6. Degenerate \evl s and the presence of symmetries.}

We now consider the case of multiple \evl s of the matrix $A_0$. This 
situation is a little bit more involved: indeed, the presence in this 
case of \coms\ $\rho\in\I_{A_0}$ of the form $\rho=v_i/v_j$ prevents the 
direct application of the argument used in the previous Theorems; another 
difficulty is related to the greater number of matrices 
$B_i\in\C(A_0)$ (if an \evl\ $\s_i$ has multiplicity $d$, then any 
$d\times d$ matrix acting on the subspace of the corresponding 
eigenvectors clearly commutes with $A_0$): this would require the 
presence of a greater number of parameters $\l$, in order to satisfy the 
Condition A.

However, the presence of degenerate \evl s is usually connected to the 
existence of some \sy\ property of the problem, and we will restrict 
to consider this simpler -- and probably more realistic and
physically interesting -- case.

We refer here to the case of ``geometric'' or Lie point-symmetries 
[18-19]: a vector function $s(u)=Lu+S(u)$ is said to be (the infinitesimal 
generator of) a \sy\ for the given DS 
if $s(u)$ is not proportional to $f(u)$ and the vector fields 
$X_f=f\cd\grad$ and $X_s=s\cd\grad$ commute:
$$[X_f,X_s]=0\en(35)$$
or, introducing the Lie-Poisson bracket $\{\cdot,\cdot\}$ between two 
vector functions $h^{(1)}(u),\ h^{(2)}(u)$ defined by
$$\{h^{(1)},h^{(2)}\}_i=\big(h^{(1)}\cdot\nabla\big)h_i^{(2)}-
\big(h^{(2)}\cdot\nabla\big)h_i^{(1)} \en(36)$$
if, equivalently,
$$\{f,s\}=0 \en(35')$$
The \sy\ is linear if $S=0$. Notice that 
linear and nonlinear \sys\ are changed one into each other under 
(nonlinear) coordinate transformations $u\to v$. It can also be remarked 
that a DS in NF always admits a linear \sy\ [8,11,12]:
\smallskip\pn
{\bf Lemma 4.} {\it Any NF admits the linear \sy\ $s_{A_0}=A_0v$:}
$$\{A_0v,g(v)\}=0$$
This is in fact a restatement of (7) and (5) using the above definition 
(36), indeed $\A(g)=\{A_0v,g\}$.
\smallskip
We need the following important properties of  Lie point-symmetries of a DS.
\pn
{\bf Lemma 5.} {\it Assume that the given DS admits a \sy\ $s(u)=Lu+S(u)$ 
where $L$ is semisimple and not zero. Then one has in particular
$$[A_0,L]=0$$  
and there is a NF of the DS which admits the linear \sy\ $s_L=Lv$.}
\smallskip\pn
The proof is well known and can be found, e.g., in [8,11,12].

An (unpleasant) consequence of the degeneracy of the \evl s $\s_i$ of 
$A_0$ is a larger arbitrarity in the choice of the matrices 
$B_i\in\C(A_0)$ to express the NF (8): consider e.g. the case in $\R^3$
$$A_0={\rm diag}(1,1,-2)$$
then the same resonant term can be written in two apparently different 
forms (the notations are obvious):
$$\eqalign{ \pmatrix{x^2yz\cr 0\cr 0}= & x^2z\pmatrix{0 & 1 & 0 \cr 0 & 0 & 0 
\cr 0  & 0 & 0}=\rho_1 B_1 v \qq {\rm or\ also}\ \cr = & 
xyz\pmatrix{ 1 & 0 & 0 \cr 0 & 0 & 0 \cr 0  & 0 & 0}=\rho_2 B_2 v
\qq \rho_1,\rho_2\in\I_{A_0},\q B_1,B_2\in\C(A_0) }$$

Assume now that the \sy\ ``removes the degeneracy'', that is, recalling 
that $A_0$ and $L$ commute, assume that there is a simultaneous basis of 
eigenvectors for $A_0$ and $L$ such that any two eigenvectors with the same 
\evl\ under $A_0$ are distinguished by a different \evl\ under $L$. Then, 
there are precisely $n$ independent matrices, which we now denote  by 
$\~B_i$, commuting with both $A_0$ and $L$:
$$ \~B_i\in\C(A_0)\cap\C(L) \en(37)$$
These matrices $\~B_i$ also commute with each other (they in fact can be 
taken diagonal: this is true upon complexification of the  space, 
in the case the \evl s of $A_0$ or of $L$ are not real: see
Example 4 below). We can then express the NF 
by means of precisely these  $n$ matrices $\~B_i$:
$$\eqalign { G(v,\l)= & \sum_{i=1}^n \b_i\(\l,\rho(v)\)\~B_iv
\qq\qq\qq \~B_i\in\C(A_0)\cap\C(L) \cr
=  & \a\(\l,\rho(v)\big)A_0v+\sp_j \b_j\big(\l,\rho(v)\)\~B_jv }\en(38)$$
using previously introduced  notations. 
The convenience of this choice is immediately evident, indeed, recalling 
Lemma 5, the NF admits the \sy\ $s_L=Lv$, i.e.
$$\{Lv,G(v)\}=0\en(39)$$
which implies
$$\sum_{i=1}^n Lv\cd\grad\b_i\ \~B_iv=0$$
or -- due to the independence of the $\~B_iv$ --
$$\a\(\l,\rho(v)\),\ \b_j\(\l,\rho(v)\)\in\I_L
\qq (j=1,\ldots,n-1) \en(40)$$
where $\I_L$  is the set of the \coms\ of the linear problem $\.v=Lv$, 
and this means that the functions $\b_i$ in (38) (and the $\rho$ as well) 
can be chosen as simultaneous \coms\ of the two linear problems 
$\.v=A_0v$ and $\.v=Lv$:
$$\rho(v),\ \b_i(\l,\rho)\in\I_{A_0}\cap\I_L \en(38')$$ 
The assumption that $L$ has removed the degeneracy has the other 
consequence that no \coms\ of the form $v_i/v_j$ are admitted in 
$\I_{A_0}\cap\I_L$ \big(and then in the NF (38)\big), therefore, 
the functions $\b_j$ appearing in the $\sp_j$ in (38) can be written 
in the form
$$\b_j(\l,\rho)=\b_j(\l,0)+\b_j^{(1)}(\l,\rho)\en(41)$$
with $\b_j^{(1)}(\l,0)=0$. On the other hand, considering the first-order 
terms (in $u$) of the DS in its initial form, and writing
$$\eqalign {\.u=& A_0u+A^{(1)}(\l)u+{\rm higher\ order\ terms} \cr
= & A_0u+\sum_ib_i(\l)\~B_iu+\sum_\ell c_\ell(\l)C_\ell u +{\rm h.o.t.} \cr
= & A_0u+a(\l)A_0u+\sp_jb_j(\l)\~B_ju+\sum_\ell c_\ell(\l)C_\ell u 
+{\rm h.o.t.}}\en(42)$$
where  $A^{(1)}(0)=0$, the sum $\sum_\ell$ includes all linear terms 
which are not resonant (i.e. $[A_0,C_\ell]\not=0$), 
and therefore disappear after the NT. The remaining terms are instead not 
changed by the NT and therefore one gets
$$a(\l)=\a(\l,0), \ b_j(\l)=\b_j(\l,0) \qq (j=1,\ldots,n-1) \en(43)$$
Then, assuming that there are $n-1$ parameters $\l_k$ and that the 
$(n-1)\times (n-1)$ matrix $\~D$, introduced by means of the following 
definition, is not singular, i.e. that
$$\det \~D\equiv\det\ {\pd b_j\over{\pd\l_k}}\Big|_{\l=0}\not=0 
\qq b_j=b_j(\l)=\b_j(\l,0) \en(44)$$
one can proceed exactly as in Theorem 1 and conclude with the 
existence of a convergent NT and of a \bif ing solution on some manifold 
$\l_j=\l_j(\rho)$.

We can then state:
\smallskip\pn
{\bf Theorem 3.} {\it Let $\.u=A(\l)u+F(\l,u)$ be a DS with $n-1$ real
 parameters 
$\l_j$; assume that $A_0$ has degenerate resonant \evl s, and that 
the DS admits 
a Lie point-symmetry generated by $s(u)=Lu+S(u)$ where $L\not=0$ is 
semisimple, such that $L$ removes the degeneracy of the \evl s of $A_0$. 
Let $\~B_i$ be $n$ independent matrices in $\C(A_0)\cap\C(L)$, and write the 
linear part (in $u$) of the DS in the form (42). Then, if the matrix $\~D$ 
defined in (44) is not singular, there is resonant \bif ing solution of the 
same form (26) as in Theorem 2.}
\smallskip
This result can be well illustrated by two examples \(notice that in the 
second one we will consider also the possibility of extending the above 
procedure in the presence of a {\it discrete} \sy\ (i.e. not a continuous 
Lie point-symmetry)\).
\pn
{\it Example 4.} Introducing the following $2\times 2$ matrices
$$J_2=\pmatrix{ & 1 \cr -1 & }\qq I_2=\pmatrix{ 1 & \cr & 1}\qq 
\en(45) $$
consider the $4-$dimensional DS, describing two coupled oscillators with 
the same unperturbed frequency $\om=1$,
$$\.u=f(u,\l)=A_0u+\sum \phi_i(r_a^2,r_b^2,r_a\times r_b,\l)M_iu \q {\rm 
with}\q A_0=\pmatrix{J_2 & \cr  & J_2 }\en(46)$$
where the \evl s of $A_0$ are $\s_i=\pm i$ (doubly degenerate), 
$r_a^2=u_1^2+u_2^2,\ r_b=u_3^2+u_4^2,\ r_a\times r_b=u_1u_4-u_2u_3$ 
and the $M_i$ are matrices of the form
$$M_i=\pmatrix{N'J_2 & N''J_2 \cr J_2N'' & J_2N' } \qq {\rm with}\ N',N'' 
\ {\rm arbitrary}\q 2\times 2 \q {\rm matrices}$$
This DS admits the linear \sy\ generated by $Lu\cd\grad$ where
$$L=\pmatrix{ &  J_2 \cr J_2 & } $$
There are $4$ matrices $\~B_i\in\C(A_0)\cap\C(L)$, namely 
$$ A_0, L,\ {\rm and}\ I=\pmatrix{I_2 & \cr & I_2}, 
H=\pmatrix{ & I_2 \cr I_2 & }$$
let us then write down explicitly the linear part (in $u$) of the DS (46) 
as in (42), i.e.
$$\eqalign { \grad f|_{u=0} = & A(\l)=A_0+A^{(1)}(\l) \cr
= & A_0+a(\l)A_0+\big(b_1(\l)I+b_2(\l)L+b_3(\l)H\big)+\sum_{\ell=1}^4 
c_\ell(\l)C_\ell }$$
where $a,b_j,c_\ell$ are combinations of the $\phi_i(0,0,0,\l)$.
When in NF, the last sum disappears, whereas the other terms remain 
unchanged, then the NF will have the form
$$\.v=A_0v+\a A_0v+\big(\b_1(\l,\rho)I+\b_2(\l,\rho)L+\b_3(\l,\rho)H\big)v
\en(47)$$
where $\rho(v)=v_1^2+v_2^2+v_3^2+v_4^2,\ v_1v_3+v_2v_4
\in\I_{A_0}\cap\I_L$, and
$$\b_j(\l,0)=b_j(\l) \qq j=1,2,3$$
The assumption (44) $\det(\pd b_j/\pd \l_k)|_{\l=0}\not=0$ then ensures, 
as already discussed, the existence 
of a resonant \bif ing solution, which in this case is a periodic 
solution with frequency $\om=1+\ldots$ and preserving the strict frequency
resonance $1:1$. \hfill$\diamondsuit$
\pn
{\it Example 5.}
As a even simpler example, let us consider the case of a $3-$dimensional 
DS with real \evl s: let
$$\.u=A_0u+\phi_1(r^2,z,\l)Lu+\phi_2(r^2,z,\l)Iu\en(48) $$
where $u\in\R^3,\ \l\in\R^2$, $r^2=u_1^2+u_2^2,\ z=u_3$ and, using 
the $2\times 2$ matrices introduced in (45), 
$$A_0=\pmatrix{I_2 &\cr & -2 } \qq L=\pmatrix{ J_2 &\cr & 0}$$
This DS admits the linear \sy\ $SO_2$ generated by $Lu\cd \grad$. 
There are 3 matrices $\~B_i$ in $\C(A_0)\cap\C(L)$, namely $A_0,L,I$ 
\big(notice that, actually, in this case, $\C(\A_0)\subset\C(L)$, but the DS 
(48) is not in NF, 
because $r^2$ and $z$ are not in $\I_{A_0}$\big). Proceeding as before,  
the NF will have the form
$$\.v=A_0v+\a A_0v+\b_1Lv+\b_2Iv \en(49)$$
where now $\b_j$ (and $\a$, of course) are functions of 
$\rho=(v_1^2+v_2^2)v_3$ and 
$\l$ only. Then we need two real parameters $\l_k$, and -- with the 
assumption (44) -- a \bif ing solution is obtained
$$\eqalign { \^u_1=&\^u_{10}\exp(1+\a)t +{\rm h.o.t.} \cr
\^u_2=&\^u_{20}\exp(1+\a)t +{\rm h.o.t.} \cr
\^u_3=&\^u_{30}\exp\big(-2(1+\a)\big)t +{\rm h.o.t.} } $$
along with some conditions
$$\l_1=\l_1(r^2z);\qq \l_2=\l_2(r^2z)$$
\hfill$\diamondsuit$
\medskip
This example can be useful to show that also the presence of {\it 
discrete} \sys\ (e.g. exchange or reflection \sys ) can be of some help 
in this approach:  a 
DS admits a discrete \sy\ $R$ , where $R$ is a nonsingular matrix, if
$$f(Ru)=Rf(u) \en(50)$$
and it is known that the NF also admits the same \sy\ [5]. The possible 
role of this fact in our argument can be illustrated by the following 
modification of the above example.
\pn
{\it Example 5 $'$.} The same as example 5, but now assume the DS admits the 
discrete \sy
$$R=\pmatrix{ 0 & 1 & \cr 1 & 0 & \cr  & & 1 } $$
Then, in both the initial DS (48) and its NF (49), the term containing the 
matrix $L$ disappears; therefore, Condition A requires the vanishing of only 
one term in the NF and a \bif ing solution can be obtained with 
the presence of just {\it one} real parameter $\l$. \hfill$\diamondsuit$

\section{7. Final remark on the role of Condition A.}

In order to stress and illustrate the relevance of the role played in our 
argument by Condition A, let us consider this final example, which is 
essentially an adjustment (in view of the present discussion) of an 
example given, with quite different purposes, in [20].
\pn
{\it Example 6.} With $u\in\R^3$ and $\l\in\R^2$, let
$$\.u=A(\l)u+\pmatrix{-(r^2+3z^2)u_1 \cr -(r^2+3z^2)u_2 \cr (3r^2+z^2)z }
\qq {\rm where} \qq A(\l)=\pmatrix{\l_1+\l_2 & 1 & 0 \cr -1 & \l_1 & 0 \cr
                                   0 & 0 & -\l_1 } \en(51)$$
with $r^2=u_1^2+u_2^2,\ z=u_3$. It is not difficult to see (cf. [20])
that, if $\l_2=0$, this system admits a family of heteroclinic orbits 
connecting the origin to the circle $r^2=\l_1$ and another family of 
heteroclinic orbits, living on the manifold $r^2+z^2=\l_1$, connecting 
the circle $r^2=\l_1$ to the point $P\equiv(0,0,\l_1)$. When $\l_2\not=0$, this 
heteroclinic structure breaks down, and an application of the Melnikov 
theory [20-23] shows the occurrence of transversal intersections of stable and 
unstable manifolds, with the consequent appearance of the chaotic 
behaviour described by the classical Birkhoff-Smale horseshoe-like 
structure (actually, we need here a simple 3-dimensional version of the 
standard Melnikov theorem: see e.g. [20]). 

On the other hand, the NF of the above DS (51) exhibits a perfectly regular 
(i.e., non-chaotic) behaviour: indeed, the NF, according to Lemma 4, must 
possess the linear symmetry generated by $A_0$, which implies that the NF 
is symmetric under rotations around the $z-$axis; then, the NF is 
essentially a 2-dimensional problem and therefore no chaos is admitted. This 
clearly implies that the  NT cannot be convergent.

If one now imposes Condition A on the NF of the DS (51), it can be easily 
seen that this condition is satisfied only along the circle defined by
$$\l_1=3(v_1^2+v_2^2)+v_3^2 \qq\qq \l_2=4(v_3-v_1^2-v_2^2) $$
where in fact a (Hopf-type) periodic solution occurs. Then, in 
conclusion, we are here in the presence of a  regular solution 
(where Condition A is indeed satisfied), which is completely surrounded by 
chaotic solutions: this clearly confirms the crucial role played by 
Condition A in the argument.

Just for completeness, and in agreement with Lemma 2 and Corollary 2, let 
us remark that this example admits (quite trivially) a reduction 
according to Remark 4: indeed, in correspondence to the \evl\ $\s_3=0$, 
we also get the stationary bifurcating solution $\l_1=z^2$ with 
$u_1=u_2=0$. \hfill$\diamondsuit$

\bigskip
\baselineskip 0.5 cm 
\Ref

[1] S.-N. Chow and J.K. Hale, ``Methods of Bifurcation Theory'', Springer, 
New York, 1982

[2] V. I. Arnold, ``Geometrical methods in the theory of differential
equations'';  Springer, Berlin, 1982

[3] V. I. Arnold and  Yu. S. Il'yashenko, ``Ordinary differential
equations''; in:  Encyclopaedia of Mathematical Sciences - vol. I,
Dynamical Systems I (D.V. Anosov and V.I. Arnold, eds.), pp. 1-148;
Springer, Berlin, 1988

[4]  A.D. Bruno:  Analytical form of differential equations, 
{\it Trans. Moscow Math. Soc.} {\bf 25} (1971), 131-198 ; and {\bf 26} 
(1972), 199-239 

[5] A.D. Bruno, ``Local methods in nonlinear differential equations'';
Springer, Berlin 1989

[6] J.-C. van der Meer, ``The Hamiltonian Hopf bifurcation'', Springer,
Berlin, 1985

[7] H.W. Broer and F. Takens,
Formally symmetric normal forms and genericity,
{\it Dynamics Reported} {\bf 2} (1989), 39-59 

[8] G. Iooss and M. Adelmeyer, ``Topics in bifurcation theory and
applications'', World Scientific, Singapore, 1992

[9] S.-N. Chow, C. Li and D. Wang, 
``Normal forms and bifurcations of planar vector fields'' 
Cambridge Univ. Press, Cambridge, 1994

[10] D. Arnal, M. Ben Ammar, and  G. Pinczon, The Poincar\'e-Dulac 
theorem for nonlinear representations of nilpotent Lie algebras,
 {\it Lett. Math. Phys.} {\bf 8} (1984), 467-476

[11] G. Cicogna and  G. Gaeta, Symmetry and Perturbation Theory in 
Nonlinear Dynamics, Springer, Berlin, 1999

[12] C.Elphick, E.Tirapegui, M.E. Brachet, P.Coullet, and G.Iooss,
A simple global characterization for normal forms of
singular vector fields, {\it Physica D} {\bf 29},  (1987) 95-127

[13] S. Walcher: On differential equations in normal form, 
{\it  Math. Ann.} {\bf 291} (1991), 293-314 

[14] G. Cicogna and  G. Gaeta, Poincar\'e normal forms and Lie point 
symmetries, {\it Journ. Phys. A: Math. Gen.}
{\bf 27} (1994), 461-476 and 7115-7124 

[15] R.H. Rand and D. Armbruster, ``Perturbation methods, bifurcation 
theory and computer algebra'', Springer, Berlin 1987

[16] G. Cicogna and G. Gaeta,  Lie point symmetries and nonlinear dynamical 
 systems, {\it Math. Comput. Modelling} {\bf 25} (1997), 101-113 

[17] G. Cicogna, Multiple-periodic bifurcation and resonance in dynamical 
systems, {\it Nuovo Cimento B} {\bf 113} (1998), 1425-1430

[18] P.J. Olver,``Applications of Lie groups to differential
equations'', Springer, Berlin, 1986

[19] L.V. Ovsjannikov, ``Group properties of differential equations'',
Novosibirsk, 1962; (English transl. by G.W. Bluman, 1967); and
``Group analysis of differential equations'', Academic
Press, New York, 1982

[20] P.Holmes, Unfolding a degenerate nonlinear oscillator: a 
  codimension two bifurcation,
{\it Ann. New York Acad. of Sciences} {\bf 357} (1980), 473-488

[21] V.K. Melnikov, Stability of the center to time periodic 
perturbation, {\it Trans. Moscow Math. Soc.} {\bf 12} (1963), 3-57

[22] S.-N. Chow, J.K. Hale and J. Mallet-Paret, An example of bifurcation 
to homoclinic orbits, {\it J. Diff. Eq.} {\bf 37} (1980), 351-373

[23] J. Guckenheimer and P.J. Holmes, ``Nonlinear oscillations, dynamical 
systems and bifurcations of vector fields'', Springer, Berlin, 1983

\bye